\documentclass[sn-nature]{sn-jnl}
\usepackage{graphicx}%
\usepackage{multirow}%
\usepackage{amsmath,amssymb,amsfonts}%
\usepackage{amsthm}%
\usepackage{mathrsfs}%
\usepackage[title]{appendix}%
\usepackage{xcolor}%
\usepackage{textcomp}%
\usepackage{manyfoot}%
\usepackage{booktabs}%
\usepackage{algorithm}%
\usepackage{algorithmicx}%
\usepackage{algpseudocode}%
\usepackage{listings}%
\usepackage{hyperref}

\usepackage{float}
\usepackage{caption}
\usepackage{geometry}
\hypersetup{
    colorlinks=true,
    linkcolor=blue,
    citecolor=blue,
    urlcolor=blue,
    pdftitle={Chemical evolution from rare isotopes in M dwarfs},
    pdfauthor={Dar{\'i}o Gonz{\'a}lez Picos, Ignas Snellen, Sam de Regt}
}

\newcommand{\Msun}{$M_{\odot}$}

\def\kms{\ifmmode{\rm km\th s^{-1}}\else km\th s$^{-1}$\fi}
\def\th{\thinspace}

\newcommand{\water}{H$_2$O}
\newcommand{\twelveCO}{\textsuperscript{12}CO}
\newcommand{\thirteenCO}{\textsuperscript{13}CO}

\newcommand{\sixteenOwater}{H$_2$\textsuperscript{16}O}
\newcommand{\CeighteenO}{C\textsuperscript{18}O}
\newcommand{\eighteenO}{\textsuperscript{18}O}
\newcommand{\eighteenOwater}{H$_2$\textsuperscript{18}O}
\newcommand{\seventeenO}{\textsuperscript{17}O}

\newcommand{\ratio}[3]{\textsuperscript{#2}#1/\textsuperscript{#3}#1}

\newcommand{\pRT}{\texttt{petitRADTRANS}}

\newcommand{\vsini}{$v\sin{i}$}

\newcommand{\Cratio}{\textsuperscript{12}C/\textsuperscript{13}C}
\newcommand{\Oratio}{\textsuperscript{16}O/\textsuperscript{18}O}

\newcommand{\fastchem}{\texttt{FastChem}}

\theoremstyle{thmstyleone}%
\theoremstyle{thmstyletwo}%

\theoremstyle{thmstylethree}%

\begin{document}

\title[Chemical evolution imprints in the rare isotopes of nearby M dwarfs
]{Chemical evolution imprints in the rare isotopes of nearby M dwarfs
}

\author[1]{\fnm{Dar{\'i}o} \sur{Gonz{\'a}lez Picos}}\email{picos@strw.leidenuniv.nl}

\author*[1]{\fnm{Ignas} \sur{Snellen}}\email{snellen@strw.leidenuniv.nl}
\author[1]{\fnm{Sam} \sur{de Regt}}\email{regt@strw.leidenuniv.nl}

\affil*[1]{\orgdiv{Leiden Observatory}, \orgname{Leiden University}, \orgaddress{\street{Einsteinweg 55}, \city{Leiden}, \postcode{2333 CC}, \country{The Netherlands}}}

\abstract{
Elements heavier than hydrogen and helium, collectively termed metals, were created inside stars and dispersed through space at the final stages of stellar evolution. The relative amounts of different isotopes (variants of the same element with different masses) in stellar atmospheres provide clues about how our galaxy evolved chemically over billions of years. M dwarfs are small, cool, long-lived stars that comprise three-quarters of all stars in our galaxy. Their spectra exhibit rich fingerprints of their composition, making them potential tracers of chemical evolution. We measure rare carbon and oxygen isotopes in 32 nearby M dwarfs spanning a range of metallicities using high-resolution infrared spectroscopy. We find that stars with higher metal content have lower \textsuperscript{12}C/\textsuperscript{13}C ratios, indicating they formed from material progressively enriched in \textsuperscript{13}C over time. This pattern is consistent with models where novae eruptions contributed significant amounts of \textsuperscript{13}C to the interstellar medium over the past few billion years. Our measurements of \textsuperscript{16}O/\textsuperscript{18}O ratio match theoretical predictions and suggest that metal-rich stars reach \textsuperscript{16}O/\textsuperscript{18}O ratios significantly lower than the Sun. These results establish M dwarfs as tracers of chemical enrichment throughout cosmic history.
}

\maketitle

\maketitle

M dwarfs are the most common stars in the Milky Way, comprising nearly 75\% of the stellar population \citep{henrySolarNeighborhoodStandard1994,reyle10ParsecSample2021}. These stars span a broad range of temperatures (2400--3900~K) and masses, with the lowest-mass M dwarfs ($M<0.3~M_\odot$) being fully convective \citep{hayashiEvolutionStarsSmall1963}. Their spectra are rich in molecular and atomic features, providing a direct window into the chemical composition of their atmospheres. Due to their extremely long main-sequence lifetimes, M dwarfs preserve the chemical signatures of their natal molecular clouds, making them potential tracers of chemical evolution across cosmic time.

Advances in high-resolution spectroscopy have enabled the detection of minor isotopologues, such as \thirteenCO{} and C\eighteenO{}, in both stellar and planetary atmospheres \citep[e.g.,][]{tsujiNearinfraredSpectroscopyDwarfs2016,zhang13COrichAtmosphereYoung2021,xuanValidationElementalIsotopic2024}. While measurements of \Cratio{} in Sun-like stars have been achieved through visible-wavelength analysis of CH and CN features \citep{botelhoCarbonIsotopicRatio2020}, such measurements remain scarce for M dwarfs. The few existing measurements of \Cratio{} in M dwarfs have been obtained through modeling of CO lines in the \textit{K}-band (near-infrared) \citep{tsujiNearinfraredSpectroscopyDwarfs2016}, with \Oratio{} measurements additionally incorporating \textit{M}-band spectra, as demonstrated for the GJ 745 AB system \citep{crossfieldUnusualIsotopicAbundances2019}.

Here, we leverage the unique properties of M dwarfs—their ubiquity, longevity, and chemically pristine atmospheres—to provide the comprehensive constraints on chemical evolution using isotopic ratios in these stars. The carbon and oxygen isotope ratios, which originate in distinct nucleosynthetic processes, offer insights into chemical enrichment pathways and provide stringent tests for Galactic Chemical Evolution (GCE) models \citep{prantzosEvolutionCarbonOxygen1996,romanoNovaNucleosynthesisGalactic2003,zhangStellarPopulationsDominated2018,
romanoEvolutionCNOElements2022}.

\begin{figure}[ht]
    \centering
    \includegraphics[width=\textwidth]{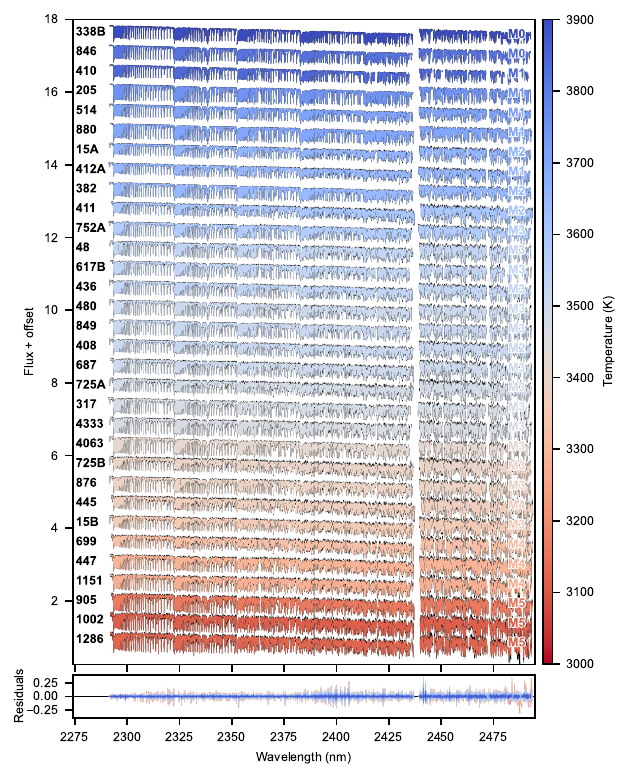}
    \caption{\textbf{Spectra of all targets in the sample.} The observed spectra (black), the best fit models (colored lines by effective temperature as shown in the colour bar) and the residuals (bottom) are shown for all 32 M dwarfs in our sample. The entire wavelength range of the spectra used in this analysis is shown.}
    \label{fig:spectra_all}
\end{figure}

We present an analysis of high-resolution ($R \sim 70{,}000$) $K$-band spectra (2.28--2.48~$\mu$m) from the SPIRou instrument \citep{donatiSPIRouNIRVelocimetry2020} for 32 nearby M dwarfs within 15~pc (see Figure \ref{fig:spectra_all}). Our sample spans effective temperatures from 3000 to 3900~K and metallicities from --0.4 to +0.4 \citep{cristofariMeasuringSmallscaleMagnetic2023}. The fundamental parameters of our sample are listed in Table \ref{tab:fundamental_parameters}. While this metallicity range is modest compared to the full extent of GCE, it corresponds to formation epochs from approximately 10~Gyr ago to the present, as inferred from empirical rotation-age relationships \citep{engleLivingRedDwarf2023,cristofariMeasuringSmallscaleMagnetic2023} and GCE models \citep{romanoGaiaESOSurveyGalactic2021,romanoEvolutionCNOElements2022}. Our sample is likely dominated by thin disk stars, with a few thick disk or halo candidates, as identified based on kinematics and $\alpha$-element abundances \citep{cristofariEstimatingFundamentalParameters2022}.

Spectra of all targets are shown in Figure \ref{fig:spectra_all} and demonstrate the high data quality and homogeneity of the sample, with typical signal-to-noise ratios exceeding 300 per spectral resolution element. Variations in spectral features are primarily driven by temperature and metallicity differences, with cooler objects exhibiting prominent \water{} lines and super-solar metallicity (i.e. [M/H] $>$ 0) stars featuring deeper absorption lines. Additionally, early spectral types are intrinsically brighter, generally resulting in higher signal-to-noise ratios.

We modelled the observed spectra using the \texttt{petitRADTRANS} radiative transfer code \citep{mollierePetitRADTRANSPythonRadiative2019}, combined with equilibrium chemistry profiles from \texttt{FastChem} \citep{kitzmannFastchemCondEquilibrium2023}. To improve the fits to the data and explore sensitivity to individual opacity sources, we allowed the abundances of chemical species to deviate from chemical equilibrium by fitting for an offset as a free parameter for each species (see Methods). The models constrain the stellar temperature structures, atmospheric compositions, and radial velocities and reproduce most spectral features in the observed wavelength range with a relative precision of 1–2\%.

\begin{figure}[ht]
    \centering
    \includegraphics[width=\textwidth]{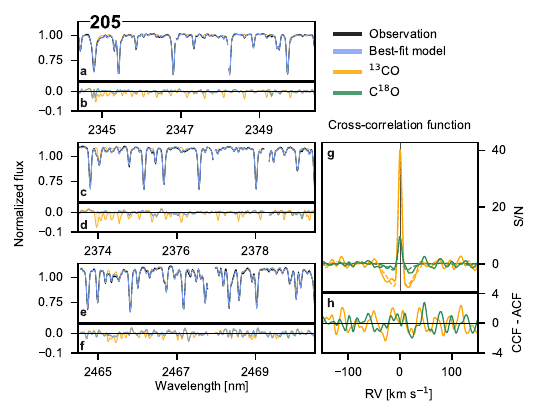}
    \caption{
        \textbf{Isotopologue detection validation.} Best-fit model spectra of Gl 205 highlighting regions with significant \thirteenCO{} and \CeighteenO{} lines. Panels \textbf{a, c, e} show observed spectra (black), fiducial best-fit model (light blue), model excluding \thirteenCO{} (orange), and model excluding \CeighteenO{} (green). Panels \textbf{b, d, f} show corresponding residuals between observed data and models. Panel \textbf{g} shows the cross-correlation function (CCF; solid line) between data and models, with auto-correlation functions (ACF; dashed lines) computed over the entire spectral range. Panel \textbf{h} shows residuals between CCF and ACF.
    }
    \label{fig:ccf}
\end{figure}

The \thirteenCO{} isotopologue is detected at greater than $2\sigma$ and greater than $3\sigma$ for 31 and 29 out of 32 targets, respectively. Similarly, \CeighteenO{} is detected at greater than $2\sigma$ and greater than $3\sigma$ for 10 and 6 stars, respectively, while upper limits were determined for the remaining targets. The robustness of these detections was assessed through Bayesian evidence comparisons between retrievals that included and excluded each isotopologue. Cross-correlation analyses further validated the reliability of these detections (see Figure \ref{fig:ccf}). By constraining the volume mixing ratios of \twelveCO{}, \thirteenCO{}, and \CeighteenO{}, we derive the carbon and oxygen isotope ratios (Table \ref{tab:isotopologue_ccf_snr}). All M dwarfs in the sample are slow rotators and exhibit narrow spectral lines, facilitating the separation of the distinct isotopologues. Spectra of all targets in our sample over the entire wavelength range used in this analysis are shown in Figure \ref{fig:spectra_all}.

The nucleosynthetic origins of carbon and oxygen isotopes determine how their observed ratios trace chemical evolution \citep{kobayashiEvolutionIsotopeRatios2011,nomotoNucleosynthesisStarsChemical2013,zhangStellarPopulationsDominated2018,
romanoEvolutionCNOElements2022}. The dominant carbon isotope, ${}^{12}$C, and the main oxygen isotope, $^{16}$O, are both synthesized in stellar helium-burning environments (triple-$\alpha$ and $\alpha$-capture reactions). In contrast, ${}^{13}$C forms as a secondary product of hydrogen burning via the CNO (carbon-nitrogen-oxygen) cycle \citep{renziniAdvancedEvolutionaryStages1981,wiescherColdHotCNO2010}. Material processed by the CNO cycle is transported to the stellar surface by convective dredge-up and ultimately expelled into the interstellar medium (ISM) through events such as supernovae or dust-driven winds \citep[][e.g.]{karakasDawesReview22014}. The second-most abundant oxygen isotope, $^{18}$O, is also produced in helium-burning shells of massive stars, but is typically destroyed by subsequent nuclear reactions. However, at low metallicity, rapid stellar rotation can induce mixing that ignites shell burning, leading to significant production of $^{18}$O and, to a lesser extent, additional ${}^{13}$C \citep{hirschiVeryLowmetallicityMassive2007,limongiPresupernovaEvolutionExplosive2018}. The yields of these isotopes depend on stellar mass, metallicity, rotation, and multiplicity \citep[e.g.,][]{romanoQuantifyingUncertaintiesChemical2005,meynetEarlyStarGenerations2006,kobayashiEvolutionIsotopeRatios2011,limongiPresupernovaEvolutionExplosive2018}. Fast-rotating massive stars in the early Galaxy are therefore thought to have produced substantial amounts of CNO isotopes, but direct isotope measurements at such low metallicities remain scarce, limiting our ability to test these predictions \citep{meynetEarlyStarGenerations2006,chiappiniNewImprintFast2008,spite12C13CRatio2021}.

\begin{figure}[h]
    \centering
    \includegraphics[width=\textwidth]{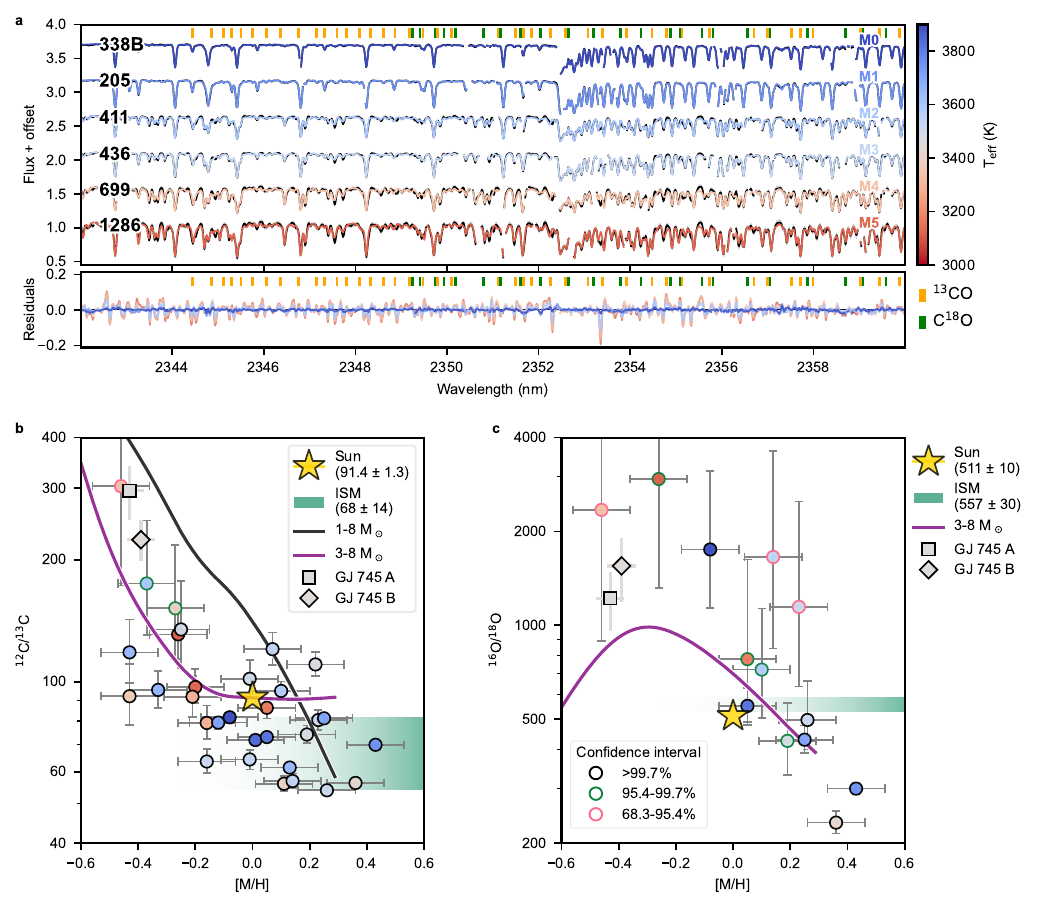}
   
    \caption{\textbf{Isotope ratios in M dwarf atmospheres.} Panel \textbf{a} shows best-fit model spectra for example targets of each spectral subtype, with observed spectra (black) and fiducial models (colored lines by effective temperature as shown in the colour bar). \thirteenCO{} and \CeighteenO{} line positions are indicated. Panels \textbf{b, c} show isotope ratios versus metallicity for our M dwarf sample (n = 32 stars). Data points are colored by effective temperature (colour bar), with edges colored by confidence intervals. Error bars show 68\% confidence intervals from Bayesian retrieval analysis for the isotope ratios and the metallicity uncertainty as reported in \cite{cristofariMeasuringSmallscaleMagnetic2023}. Purple and black lines represent GCE model tracks for nova WD progenitors (1--8 \Msun{} and 3--8 \Msun{}) \citep{romanoGaiaESOSurveyGalactic2021,romanoEvolutionCNOElements2022}. \thirteenCO{} is detected at $>3\sigma$ in 29/32 stars, \CeighteenO{} in 6/32 stars (see Table \ref{tab:isotopologue_ccf_snr}). Reference values: Sun \citep{lyonsLightCarbonIsotope2018,ayresSUNLIGHTEREARTH2013}, ISM \citep{milam1213Isotope2005,wilsonIsotopesInterstellarMedium1999}, and GJ 745 AB \citep{crossfieldUnusualIsotopicAbundances2019}.}
    \label{fig:isotope_ratios_metallicity}
\end{figure}

The bottom panels of Figure \ref{fig:isotope_ratios_metallicity} show the carbon and oxygen isotope ratios as a function of metallicity for the objects in our sample. Both isotope ratios exhibit a decrease with metallicity as predicted by some GCE models, assuming nova progenitors in the mass ranges of 1--8 \Msun{} and 3--8 \Msun{}, which have been proposed by some authors \citep{romanoEvolutionCNOElements2022} as a dominant source of ${}^{13}$C enrichment in the latest stages of GCE. Over recurrent thermonuclear explosions on the surfaces of accreting white dwarfs (WDs) the ISM is enriched with significant amounts of ${}^{13}$C. Both models predict a decreasing ${}^{12}\text{C}/^{13}\text{C}$ ratio, consistent with the data, but the 3--8 \Msun{} model exhibits a notable flattening from around solar metallicity to super-solar metallicities. This flattening arises from the interplay between the assumed star formation history in the solar neighborhood, which has significantly declined over the past 4--5 Gyr, and the lifetimes of stars with $M > 3$ \Msun{}, which are less than 500 Myr \citep{romanoQuantifyingUncertaintiesChemical2005}. When contribution from lower mass nova progenitors (1-3 \Msun{}) is included, the enrichment of ${}^{13}$C continues over recent times, decreasing the isotope ratio further.

Super-solar metallicity M dwarfs in our sample exhibit $^{12}\text{C}/^{13}\text{C}$ ratios consistent with the present-day local ISM value ($68\pm15$; \cite{milam1213Isotope2005}). Nevertheless, these ratios display substantial scatter that exceeds observational uncertainties. This dispersion likely reflects a combination of factors, including variations in local stellar formation environments, radial migration effects \citep{kubrykRadialMigrationBardominated2013}, the presence of multiple stellar populations within the sample with distinctly different chemical histories \citep{fuhrmannLocalStellarPopulations2017} and possibly underestimated uncertainties in the metallicity measurements \citep{mannHOWCONSTRAINYOUR2015}. In particular, our own Sun is thought to have migrated outward by at least 1 kpc from its birth location, as inferred from its metallicity, carbon isotope ratio, and comparison with nearby stars \citep[e.g.,][]{milam1213Isotope2005}.

\begin{figure}[h]
    \centering
    \includegraphics[width=0.66\textwidth]{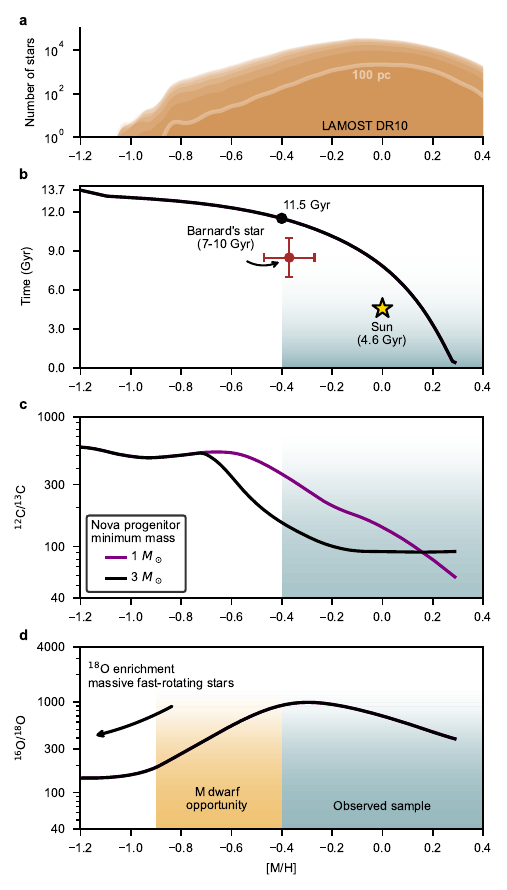}
    \caption{
        \textbf{The M dwarf opportunity for Galactic Chemical Evolution.}
        Panel \textbf{a} shows the distribution of M dwarfs within 1000~pc from the LAMOST spectroscopic survey data release 10 \citep{wangLAMOSTRevealsLonglived2025}, with the color shades  indicating distance scales in steps of 100~pc.
        Panel \textbf{b} shows the age–metallicity relation from GCE models \citep{romanoGaiaESOSurveyGalactic2021,romanoEvolutionCNOElements2022}, with individual stellar data points for Barnard's star (age: $7-10$ Gyr; \cite{ribasCandidateSuperEarthPlanet2018}; metallicity: \cite{cristofariMeasuringSmallscaleMagnetic2023}) and the Sun (age: \cite{bouvierAgeSolarSystem2010}; metallicity: \cite{asplundChemicalMakeupSun2021}).
        Panels \textbf{c, d} show evolutionary tracks of the \Cratio{} and \Oratio{} isotope ratios from GCE models as a function of [M/H], with purple and black lines representing nova WD progenitors (1--8~\Msun{} and 3--8~\Msun{}, respectively). Shaded regions indicate our sample range (gray; n = 32 stars) and proposed low-metallicity range (orange).
    }
    \label{fig:gce_models_time}
\end{figure}

The imprint of fast-rotating, massive, extremely-low-metallicity stars is proposed to be present in the oxygen isotope ratio \ratio{O}{16}{18} in stars that formed at later epochs (but still relatively low metallicity) as they are expected to inherit the low \ratio{O}{16}{18} from the ISM at the time of their formation \citep{hegerNucleosyntheticSignaturePopulation2002,meynetEarlyStarGenerations2006}. Although these low-metallicity stars are beyond the range of our current sample, enrichment signatures may still be detectable in the evolution of the \Oratio{} as a function of metallicity (see Figure \ref{fig:gce_models_time}). Specifically, state-of-the-art GCE models predict a turnover in the $^{16}\text{O}/^{18}\text{O}$ ratio near [M/H] $\sim$ –0.3, as a result of the different contributions of rotating and non-rotating stars that act on different timescales \citep{romanoGaiaESOSurveyGalactic2021,romanoEvolutionCNOElements2022}. Observations presented in Figure \ref{fig:isotope_ratios_metallicity} (panel c) tentatively support this scenario, but additional measurements are needed to further assess this.

In the most recent stages of evolution, corresponding to super-solar metallicity stars (see Figure \ref{fig:gce_models_time}b), targets of our sample at [M/H] $\sim$ 0.25 show remarkable agreement with the highest metallicity range probed by the GCE models. Extending these models to even higher metallicities is not straightforward \citep{romanoNovaNucleosynthesisGalactic2003}, but our findings motivate such efforts: the most metal-rich stars in our sample exhibit $^{16}\text{O}/^{18}\text{O}$ ratios as low as 200--300, significantly below the ISM and solar values \citep{wilsonIsotopesInterstellarMedium1999,ayresSUNLIGHTEREARTH2013}.

Studying the carbon and oxygen abundance ratios in very metal-poor M dwarfs presents both a significant challenge and an exciting opportunity. While measurements of $^{12}\text{C}/^{13}\text{C}$ exist for carbon-enhanced metal-poor (CEMP) stars—objects with extremely low iron content yet enhanced carbon abundances \citep{molaro12C13CIsotopic2023}—such data remain scarce for M dwarfs. These CEMP stars, often located near the main-sequence turnoff, are thought to be very old and have exhibited $^{12}\text{C}/^{13}\text{C}$ ratios as low as 10. However, these values may reflect contamination from AGB companions \citep{ryanOriginsTwoClasses2005}.

Notably, \cite{spite12C13CRatio2021} reported a low carbon isotope ratio of $33^{+12}_{-6}$ for a classic metal-poor star ([Fe/H] = –2.6) with no signs of internal mixing, suggesting that low $^{12}\text{C}/^{13}\text{C}$ values can be intrinsic. Extending such isotopic studies to M dwarfs would provide valuable tracers of both present-day and primordial nucleosynthetic signatures, especially at the lowest and highest metallicities.

Future efforts to expand the target sample and improve the precision of isotope ratio measurements would benefit from \textit{M}-band observations accessing the fundamental CO band. Instruments such as METIS on the ELT \citep{brandlMETISMidinfraredELT2021} and NIRSpec aboard JWST \citep{jakobsenNearInfraredSpectrographNIRSpec2022,gardnerJamesWebbSpace2023} offer a promising avenue to bridge the gap to the lowest metallicity stars and map the chemical enrichment history of the Milky Way back to its earliest epochs.

\newpage

\newpage

\clearpage

\section*{Methods}\label{sec:methods}

\textbf{Sample}  
Our sample comprises 32 nearby M dwarfs (spectral types M0–M5) with effective temperatures between 3000 and 3900~K, selected from a high signal-to-noise catalogue with robust metallicity determinations \citep{cristofariEstimatingFundamentalParameters2022, cristofariEstimatingAtmosphericProperties2022, cristofariConstrainingAtmosphericParameters2023}. The metallicity range ([M/H] = –0.4 to +0.4) corresponds to formation epochs from approximately 10~Gyr ago to the present, as estimated from empirical rotation-age relationships \citep{engleLivingRedDwarf2023,cristofariMeasuringSmallscaleMagnetic2023} and the metallicity-age relation in GCE models \citep{romanoGaiaESOSurveyGalactic2021,romanoEvolutionCNOElements2022} (see Figure \ref{fig:gce_models_time}b). Most targets are likely thin disk members, with a few possible thick disk or halo stars based on kinematics and $\alpha$-element abundances \citep{cristofariEstimatingFundamentalParameters2022}. 

\textbf{Observations}  
We employed archival high-resolution ($R \sim 70{,}000$) $K$-band spectra (2.28–2.48~$\mu$m) from the SPIRou instrument \citep{donatiSPIRouNIRVelocimetry2020}. Reduced spectra were retrieved from the Canadian Astronomy Data Centre that were automatically processed with the SPIRou reduction pipeline (\texttt{APERO} v0.7.28, \cite{cookAPEROPipelinEReduce2022}), including wavelength calibration and telluric correction. For each star, we selected only individual spectra with S/N~$>$~50, airmass~$<$~1.4, and seeing~$<$~1" to ensure high data quality. Multiple epochs were combined for each target after barycentric correction and spectral alignment, focusing on the three reddest SPIRou spectral orders, where the CO absorption features are present. The final combined spectra reach S/N of 500–1800 at 2.3~$\mu$m, with uncertainties estimated from a combination of propagated errors and the pixel-wise standard deviation across epochs. This approach robustly mitigates telluric contamination by leveraging the range of barycentric velocities and assigns larger uncertainties to variable pixels, minimizing systematic effects from coadding multi-epoch data. 

\textbf{Atmospheric models}  
High-resolution model spectra were generated using the line-by-line radiative transfer code \pRT{} \citep{mollierePetitRADTRANSPythonRadiative2019}. Molecular and atomic opacities were calculated from cross-sections based on transition data, obtained from the ExoMol and HITRAN databases for molecular species \citep{tennyson2024ReleaseExoMol2024, rothmanHITEMPHightemperatureMolecular2010}, and from the Kurucz database for atomic species \citep{kuruczNewOpacityCalculations1991}, supplemented with energy levels from the NIST Atomic Spectra Database \citep{AtomicSpectraDatabase2009}.

\textbf{Temperature-pressure profile}  
The atmospheric temperature profile was parameterised with gradients defined at seven pressure levels, following an adaptation of \cite{zhangELementalAbundancesPlanets2023a} as described in \cite{gonzalezpicosESOSupJupSurvey2024a}:
\[
    \begin{aligned}
        P_0 &= 10^2 \text{ bar}, \\  
        P_1 &= P_{\text{RCE}} - 2 \Delta P_{\text{bot}}, \\  
        P_2 &= P_{\text{RCE}} - \Delta P_{\text{bot}}, \\  
        P_3 &= P_{\text{RCE}}, \\  
        P_4 &= P_{\text{RCE}} + \Delta P_{\text{top}}, \\  
        P_5 &= P_{\text{RCE}} + 2 \Delta P_{\text{top}}, \\  
        P_6 &= 10^{-5} \text{ bar},  
    \end{aligned}
\]
where \( P_{\text{RCE}} \) is the radiative-convective boundary pressure, and \( \Delta P_{\text{bot}} \) and \( \Delta P_{\text{top}} \) are the pressure spacings below and above \( P_{\text{RCE}} \), respectively. Temperatures at intermediate atmospheric layers were linearly interpolated between these levels, resulting in 11 free parameters for the pressure-temperature profile. This approach balances flexibility with minimal assumptions about the atmospheric structure.

\textbf{Chemical abundances}  
Chemical abundances were calculated as a function of altitude using thermochemical equilibrium from \fastchem{} \citep{kitzmannFastchemCondEquilibrium2023}. Deviations from equilibrium chemistry were modeled with logarithmic offset terms:
\[
\log X_i = \log X_{i,\rm eq} + \alpha(X_i),
\]
where \( \alpha(X_i) \) is the offset for species \( i \). Abundances were interpolated from a precomputed \fastchem{} grid at each model evaluation of the temperature-pressure profile, using \texttt{scipy.interpolate.LinearNDInterpolator}. For species not included in \fastchem{} (e.g., Sc), constant volume-mixing ratios with altitude were assumed, and \( \log X_i \) was treated as a free parameter. Minor isotopologue abundances were derived from their major isotopologues using isotope ratios, ensuring consistency across altitude profiles.

\textbf{Spectral broadening}  
Rotational broadening was applied by convolving the model spectrum with a rotational profile generated using \texttt{PyAstronomy} \citep{czeslaPyAPythonAstronomyrelated2019}, with the projected rotational velocity (\vsini) treated as a free parameter. We adopted a fixed linear limb-darkening coefficient of 0.20, derived from PHOENIX models via \texttt{Exo-TiC-LD} \citep{husserNewExtensiveLibrary2013, grantExoTiCLDThirtySeconds2024}. Instrumental broadening was modeled as a Gaussian profile with a full width at half maximum corresponding to the instrument's resolving power (4.3~km/s), consistent with previous SPIRou analyses \citep{donatiSPIRouNIRVelocimetry2020, cristofariEstimatingFundamentalParameters2022}. Microturbulence was not explicitly included, as it is expected to be small for M dwarfs ($<1$~km/s; e.g., \cite{hauschildtNextGenModelAtmosphere1999}), and any additional broadening would be absorbed in the $v\sin i$ parameter \citep{hahlinMultiscaleMagneticField2024}.

\textbf{Continuum modelling}  
To account for low-frequency noise, we modeled the continuum using a linear spline with 25 knots per spectral order, following \cite{ruffioDetectingExomoonsRadial2023} and \cite{gonzalezpicosESOSupJupSurvey2024a}. This approach acts as a high-pass filter, ensuring robustness in regions with dense absorption lines and minimizing sensitivity to hyperparameter choices of the filtering process.

\textbf{Retrieval analysis}  
We used the Nested Sampling algorithm from \texttt{PyMultiNest} \citep{ferozImportanceNestedSampling2019,buchnerPyMultiNestPythonInterface2016} to fit the data and obtain posterior probability distributions for all free parameters. For each spectral order $i$, the likelihood function is:
\begin{equation}
    \ln \mathcal{L}_{i} = -\frac{1}{2} \left( N_k \ln(2\pi) + \ln|\Sigma_{0,i}| + N_k \ln(s_{i}^2) + \frac{1}{s_{i}^2} \mathbf{r}_{i}^T \Sigma_{0,i}^{-1} \mathbf{r}_{i} \right),
\end{equation}
where $N_k$ is the number of pixels, $\Sigma_{0,i}$ is the diagonal covariance matrix, $s_{i}$ is the uncertainty scaling factor, and $\mathbf{r}_{i} = \mathbf{d}_{i} - \phi_i M_i$ is the residual between the data ($\mathbf{d}_{i}$) and the model ($M_i$), with $\phi_i$ as the spline coefficients. The total likelihood is the sum over all orders: $\ln \mathcal{L} = \sum_i \ln \mathcal{L}_{i}$.

The optimal spline coefficients $\phi$ and scaling factor $s$ were determined by solving the least-squares problem:
\[
\min_{\phi,s} \sum_i \left(\mathbf{d}_{i} - \phi_i M_i\right)^T \Sigma_{0,i}^{-1} \left(\mathbf{d}_{i} - \phi_i M_i\right),
\]
using the \texttt{scipy.optimize} Non-Negative Least Squares algorithm \citep{lawsonSolvingLeastSquares1995}.

Our fiducial model includes 29 free parameters, covering stellar properties (surface gravity, radial velocity, \vsini), the temperature profile (surface temperature, gradients, pressure levels), and atmospheric composition (offsets from equilibrium chemistry or volume-mixing ratios, and isotope ratios for molecules such as \twelveCO/\thirteenCO{} and \sixteenOwater/\eighteenOwater{}).

To assess the significance of isotopologue detections, we compute the Bayes factor between the fiducial model (all species included) and a model excluding the isotopologue of interest:
\begin{align}
    \ln B = \ln \mathcal{Z}_{\text{full}} - \ln \mathcal{Z}_{\text{no iso}},
\end{align}
where $\mathcal{Z}$ is the Bayesian evidence from Nested Sampling. We interpret $\ln B > 3$ and $\ln B > 11$ as moderate and strong evidence, respectively, for the presence of the isotopologue \citep{kassBayesFactors1995}, corresponding to $3\sigma$ and $5\sigma$ detection significances \citep{bennekeHOWDISTINGUISHCLOUDY2013}.

Additionally, we perform cross-correlation analysis to further validate isotopologue detections (see Figure \ref{fig:ccf}). The cross-correlation function (CCF) is computed for the residuals between models with and without the isotopologue (e.g., \cite{zhang13COrichAtmosphereYoung2021}). The CCF signal-to-noise (S/N) is defined as the ratio of its peak value to the standard deviation in a region away from the peak (RV $> 100$~km~s$^{-1}$) of the residuals between the CCF and the autocorrelation function \citep{regtESOSupJupSurvey2024}.

\textbf{Retrieval results}  
We detect \twelveCO{}, \sixteenOwater{}, Na, Ca, Sc, and HF in all targets. The \water{} abundances vary across the sample due to thermal dissociation. OH is detected in the early M dwarfs, but not significantly in the late M dwarfs. The presence of substantial \water{} in late M dwarfs contributes to increased scatter in the residuals (see Figure \ref{fig:spectra_all}). Several spectral lines show a notable mismatch, which we identify as \water{} lines. We measure line offsets of approximately 4–5 km/s between the best fit models and the observed data for certain water lines. These shifts are both positive and negative, indicating they are not due to a systematic offset in the data. The depth of these lines varies significantly, ranging from about 2\% in the hottest targets to around 20\% in the coolest ones, reflecting the decrease in water abundance with increasing effective temperature. Importantly, these discrepancies in the position of certain \water{} lines do not affect the derived isotope ratios from CO lines, as the isotopologues are spectrally resolved and the CO lines are both more abundant and more prominent than the \water{} lines. We found that the reported values of the isotope ratios are robust against slight variations in the prior ranges of the temperature gradients.

Table \ref{tab:isotopologue_ccf_snr} presents the carbon and oxygen isotope ratios derived from the CO isotopologue measurements. While we included C\seventeenO{} in our retrievals, only lower limits were obtained across the sample. H$_2^{18}$O was weakly detected in some objects but did not yield strong detections. It is worth noting that the linelist for H$_2^{18}$O is valid only up to 3000 K and is less complete compared to the main isotopologue, H$_2^{16}$O \citep{polyanskyExoMolMolecularLine2017}.

\textbf{Data availability}.  
 The reduced SPIRou data is available at the Canadian Astronomy Data Center (CADC) at \url{https://www.cadc-ccda.hia-iha.nrc-cnrc.gc.ca/en/}. The reduced data used in this work and the derived best-fit model specta are publicly available at \citep{gonzalezpicosChemicalEvolutionImprints2025}.

\textbf{Code availability}.  
The software to calculate the cross-sections used in this work is available at \url{https://github.com/samderegt/pyROX}. The radiative transfer code to generate atmospheric models is available at \url{https://petitradtrans.readthedocs.io/en/2.7.7/}. The equilibrium chemistry code \fastchem{} is available at \url{https://github.com/NewStrangeWorlds/FastChem}. The implementation of the Nested Sampling algorithm used in this work can be found at \url{https://github.com/JohannesBuchner/PyMultiNest}.

\textbf{Correspondence and requests for materials} should be addressed to I.S. (snellen@strw.leidenuniv.nl).

\textbf{Acknowledgements}. D.G.P. thanks D. Romano for providing the GCE models and valuable insights on the evolution of isotope ratios. We thank the anonymous referees for their helpful comments and suggestions. D.G.P, I.S and S.d.R acknowledge NWO grant OCENW.M.21.010. This work used the Dutch national e-infrastructure with the support of the SURF Cooperative using grant no. EINF-4556. Based on observations obtained at the Canada-France-Hawaii Telescope (CFHT) which is operated from the summit of Maunakea by the National Research Council of Canada, the Institut National des Sciences de l'Univers of the Centre National de la Recherche Scientifique of France, and the University of Hawaii. The observations at the Canada-France-Hawaii Telescope were performed with care and respect from the summit of Maunakea which is a significant cultural and historic site. Based on observations obtained with SPIRou, an international project led by Institut de Recherche en Astrophysique et Plan{\'e}tologie, Toulouse, France.

\textbf{Author contributions} D.G.P led the data processing, analysis and wrote the manuscript. I.S contributed to the model fitting, interpretation of the results and writing of the manuscript. S.d.R contributed to the code development of the models, provided the code for the opacity calculator and advised on model fitting.  All authors contributed to the text and figures of the manuscript.

\textbf{Competing interests}
The authors declare no competing interests.

\begin{table*}[ht]
    \renewcommand{\arraystretch}{1.3}
    \centering
    \caption{Fundamental parameters of the stars in the sample.}\label{tab:fundamental_parameters}
    \begin{tabular}{lccccccc}\hline
    Star & Spectral Type & Distance / pc & M/M$_{\odot}$ & T$_{\rm eff}$ / K & [M/H] dex\\
    \hline
    Gl 338B & M0.0 & $6.33$ & $0.58 \pm 0.02$ & $3899 \pm 30$ & $-0.11 \pm 0.10$ \\
    Gl 846 & M0.5 & $10.57$ & $0.57 \pm 0.02$ & $3821 \pm 30$ & $+0.05 \pm 0.10$ \\
    Gl 410 & M1.0 & $11.94$ & $0.55 \pm 0.02$ & $3842 \pm 31$ & $-0.02 \pm 0.10$ \\
    Gl 205 & M1.5 & $5.70$ & $0.58 \pm 0.02$ & $3747 \pm 31$ & $+0.41 \pm 0.10$ \\
    Gl 514 & M1.0 & $7.63$ & $0.50 \pm 0.02$ & $3710 \pm 31$ & $-0.12 \pm 0.10$ \\
    Gl 880 & M1.5 & $6.87$ & $0.55 \pm 0.02$ & $3702 \pm 30$ & $+0.24 \pm 0.10$ \\
    Gl 15A & M2.0 & $3.56$ & $0.39 \pm 0.02$ & $3660 \pm 31$ & $-0.29 \pm 0.10$ \\
    Gl 412A & M1.0 & $4.90$ & $0.39 \pm 0.02$ & $3650 \pm 30$ & $-0.40 \pm 0.10$ \\
    Gl 382 & M2.0 & $7.71$ & $0.51 \pm 0.02$ & $3645 \pm 31$ & $+0.13 \pm 0.10$ \\
    Gl 411 & M2.0 & $2.55$ & $0.39 \pm 0.02$ & $3601 \pm 30$ & $-0.32 \pm 0.10$ \\
    Gl 752A & M3.0 & $5.91$ & $0.47 \pm 0.02$ & $3579 \pm 31$ & $+0.11 \pm 0.10$ \\
    Gl 48 & M3.0 & $8.23$ & $0.43 \pm 0.02$ & $3529 \pm 31$ & $+0.08 \pm 0.10$ \\
    Gl 617B & M3.0 & $10.77$ & $0.45 \pm 0.02$ & $3532 \pm 30$ & $+0.14 \pm 0.10$ \\
    Gl 436 & M3.0 & $9.76$ & $0.32 \pm 0.00$ & $3531 \pm 30$ & $-0.00 \pm 0.10$ \\
    Gl 480 & M3.5 & $14.26$ & $0.45 \pm 0.02$ & $3505 \pm 30$ & $+0.20 \pm 0.10$ \\
    Gl 849 & M3.5 & $8.80$ & $0.35 \pm 0.00$ & $3512 \pm 30$ & $+0.26 \pm 0.10$ \\
    Gl 408 & M2.5 & $6.75$ & $0.38 \pm 0.02$ & $3512 \pm 31$ & $-0.13 \pm 0.10$ \\
    Gl 687 & M3.0 & $4.55$ & $0.41 \pm 0.04$ & $3498 \pm 30$ & $+0.02 \pm 0.10$ \\
    Gl 725A & M3.0 & $3.52$ & $0.26 \pm 0.00$ & $3485 \pm 31$ & $-0.21 \pm 0.10$ \\
    Gl 317 & M3.5 & $15.18$ & $0.42 \pm 0.02$ & $3473 \pm 31$ & $+0.23 \pm 0.10$ \\
    Gl 4333 & M3.5 & $10.60$ & $0.37 \pm 0.02$ & $3467 \pm 31$ & $+0.26 \pm 0.10$ \\
    Gl 4063 & M3.5 & $10.89$ & -- & $3419 \pm 31$ & $+0.42 \pm 0.10$ \\
    Gl 725B & M3.5 & $3.52$ & $0.21 \pm 0.00$ & $3399 \pm 30$ & $-0.21 \pm 0.10$ \\
    Gl 876 & M3.5 & $4.67$ & $0.33 \pm 0.00$ & $3395 \pm 30$ & $+0.16 \pm 0.10$ \\
    Gl 445 & M4.0 & $5.25$ & $0.24 \pm 0.02$ & $3379 \pm 31$ & $-0.17 \pm 0.10$ \\
    Gl 15B & M3.5 & $3.56$ & $0.15 \pm 0.00$ & $3362 \pm 30$ & $-0.33 \pm 0.10$ \\
    Gl 699 & M4.0 & $1.83$ & $0.15 \pm 0.00$ & $3326 \pm 31$ & $-0.37 \pm 0.10$ \\
    Gl 447 & M4.0 & $3.37$ & $0.18 \pm 0.02$ & $3291 \pm 30$ & $-0.07 \pm 0.10$ \\
    Gl 1151 & M4.5 & $8.04$ & $0.17 \pm 0.02$ & $3278 \pm 31$ & $-0.03 \pm 0.10$ \\
    Gl 905 & M5.0 & $3.15$ & $0.14 \pm 0.00$ & $3161 \pm 31$ & $+0.24 \pm 0.10$ \\
    Gl 1002 & M5.5 & $4.85$ & $0.12 \pm 0.02$ & $3110 \pm 32$ & $-0.03 \pm 0.10$ \\
    Gl 1286 & M5.5 & $7.18$ & $0.12 \pm 0.02$ & $3121 \pm 31$ & $+0.08 \pm 0.10$ \\
    \hline
    \end{tabular}
    \\
    \footnotesize{\textbf{Notes:} The spectral types, effective temperatures and masses are from \citep{cristofariEstimatingAtmosphericProperties2022}. Distances are from Gaia EDR3, with typical uncertainties of 0.002 pc \citep{gaiacollaborationGaiaEarlyData2021}. Metallicities were measured on the same dataset as the one used for the present work by \cite{cristofariMeasuringSmallscaleMagnetic2023}.}
\end{table*}

\begin{table}
    \renewcommand{\arraystretch}{1.3}
    \caption{Results and detection significances of the CO isotopologues.}
    \label{tab:isotopologue_ccf_snr}
    \begin{tabular*}{0.85\textwidth}{lcccccccccc}
    \toprule
    & \multicolumn{4}{@{}c@{}}{\rule[0.0ex]{40pt}{0pt}$^{13}$CO}& \multicolumn{6}{@{}c@{}}{\rule[0.0ex]{44pt}{0pt}C$^{18}$O} \\\cmidrule{3-6}\cmidrule{8-11}
    Gl & &$\log{} ^{12}$C/$^{13}$C& $\ln B^{\rm a}$ & $\sigma$ & CCF$^{\rm b}$ & &$\log{} ^{16}$O/$^{18}$O& $\ln B^{\rm a}$ & $\sigma$ & CCF$^{\rm b}$\\
    \cmidrule{1-1}\cmidrule{3-6}\cmidrule{8-11}
    338B & &$1.91^{+0.01}_{-0.01}$ & 482.4 & $> 10$ &16.7 &  &$3.24^{+0.25}_{-0.19}$ & 5.9 & 3.9 &--\\
    846 & &$1.86^{+0.01}_{-0.01}$ & -938.6 & $> 10$ &23.1 &  &$2.74^{+0.06}_{-0.05}$ & 17.6 & 6.2 &3.1\\
    410 & &$1.86^{+0.01}_{-0.01}$ & -354.1 & $> 10$ &12.5 &  &$ >2.92$ & -5.4 & -- &--\\
    205 & &$1.84^{+0.01}_{-0.01}$ & 2154.1 & $> 10$ &40.4 &  &$2.48^{+0.02}_{-0.02}$ & 176.9 & $> 10$ &9.6\\
    514 & &$1.90^{+0.02}_{-0.02}$ & 165.4 & $> 10$ &9.3 &  &$ >3.03$ & -20.1 & -- &--\\
    880 & &$1.91^{+0.01}_{-0.01}$ & 817.0 & $> 10$ &20.5 &  &$2.61^{+0.04}_{-0.04}$ & 35.3 & 9.0 &3.7\\
    15A & &$1.98^{+0.05}_{-0.05}$ & 19.4 & 6.5 &3.8 &  &$ >3.22$ & -3.4 & -- &--\\
    412A & &$2.07^{+0.08}_{-0.08}$ & 5.4 & 3.7 &-- &  &$ >3.05$ & -3.8 & -- &--\\
    382 & &$1.79^{+0.01}_{-0.01}$ & 561.7 & $> 10$ &16.8 &  &$ >2.74$ & -3.0 & -- &--\\
    411 & &$ >2.11$ & 1.9 & 2.5 &-- &  &$ >3.16$ & -2.0 & -- &--\\
    752A & &$1.98^{+0.02}_{-0.02}$ & 117.0 & $> 10$ &7.3 &  &$ >2.70$ & 1.9 & 2.5 &--\\
    48 & &$2.08^{+0.04}_{-0.04}$ & 23.4 & 7.1 &3.8 &  &$ >3.27$ & -0.7 & -- &--\\
    617B & &$1.76^{+0.02}_{-0.02}$ & 195.8 & $> 10$ &10.1 &  &$ >2.93$ & 0.3 & 1.5 &--\\
    436 & &$1.81^{+0.02}_{-0.02}$ & 105.2 & $> 10$ &7.4 &  &$ >2.90$ & -1.2 & -- &--\\
    480 & &$1.91^{+0.02}_{-0.02}$ & 99.3 & $> 10$ &7.3 &  &$ >2.80$ & 0.1 & 1.3 &--\\
    849 & &$1.73^{+0.01}_{-0.01}$ & 459.2 & $> 10$ &14.8 &  &$2.69^{+0.13}_{-0.10}$ & 4.9 & 3.6 &--\\
    408 & &$1.80^{+0.03}_{-0.03}$ & 55.6 & $> 10$ &5.3 &  &$ >3.14$ & -6.9 & -- &--\\
    687 & &$2.01^{+0.05}_{-0.04}$ & 22.7 & 7.0 &3.8 &  &$ >3.19$ & -2.3 & -- &--\\
    725A & &$2.13^{+0.12}_{-0.10}$ & 4.8 & 3.6 &-- &  &$ >3.24$ & -2.0 & -- &--\\
    317 & &$1.87^{+0.02}_{-0.02}$ & 126.7 & $> 10$ &8.3 &  &$ >2.52$ & 3.0 & 2.9 &--\\
    4333 & &$2.04^{+0.03}_{-0.03}$ & 57.7 & $> 10$ &5.7 &  &$ >2.74$ & -0.0 & -- &--\\
    4063 & &$1.75^{+0.01}_{-0.01}$ & 572.3 & $> 10$ &18.7 &  &$2.37^{+0.04}_{-0.04}$ & 38.2 & 9.3 &4.2\\
    725B & &$ >2.06$ & 1.9 & 2.5 &-- &  &$ >3.15$ & -1.2 & -- &--\\
    876 & &$1.75^{+0.02}_{-0.02}$ & 177.4 & $> 10$ &10.6 &  &$ >2.68$ & -0.4 & -- &--\\
    445 & &$ >2.12$ & -1.0 & -- &-- &  &$ >3.16$ & -1.7 & -- &--\\
    15B & &$1.96^{+0.08}_{-0.07}$ & 5.8 & 3.8 &-- &  &$ >3.08$ & -5.8 & -- &--\\
    699 & &$ >2.24$ & 0.8 & 1.9 &-- &  &$ >2.95$ & 0.3 & 1.5 &--\\
    447 & &$1.96^{+0.05}_{-0.05}$ & 20.8 & 6.7 &3.7 &  &$ >3.15$ & -3.4 & -- &--\\
    1151 & &$1.90^{+0.04}_{-0.04}$ & 32.8 & 8.7 &4.6 &  &$ >3.28$ & -1.4 & -- &--\\
    905 & &$1.94^{+0.03}_{-0.03}$ & 93.5 & $> 10$ &7.4 &  &$ >2.68$ & 1.0 & 2.0 &--\\
    1002 & &$2.12^{+0.07}_{-0.06}$ & 11.3 & 5.1 &3.2 &  &$ >3.12$ & 1.3 & 2.2 &--\\
    1286 & &$1.99^{+0.04}_{-0.04}$ & 31.9 & 8.3 &4.8 &  &$ >3.31$ & -2.5 & -- &--\\
    \botrule
    \end{tabular*}
    
    \footnotesize{$^{\rm a}$ Logarithm of the Bayes factor between the fiducial model and a model without the given species.}
    
    \footnotesize{$^{\rm b}$ Signal-to-noise ratio of the cross-correlation function. Missing values denote non-detections.}
\end{table}

\end{document}